\documentclass[english,reprint, aip,jcp,superscriptaddress,showpacs,floatfix]{revtex4-2}
\usepackage{amsmath}
\usepackage{graphicx, subfigure}
\usepackage{makeidx}

\begin{document}

\title{Colloidal logic-gate circuits can process environmental signals and autonomously perform tasks}
\author{Jiang-Xing Chen}
\altaffiliation{jxchen@hdu.edu.cn}
\affiliation{Department of Physics, Hangzhou Normal University, Hangzhou 311121, China}
\author{Jia-Qi Hu}
\affiliation{Department of Physics, Hangzhou Normal University, Hangzhou 311121, China}
\author{Raymond Kapral}
\altaffiliation{rkapral@chem.utoronto.ca}
\affiliation{Chemical Physics Theory Group, Department of Chemistry, University of Toronto, Toronto, Ontario M5S 3H6, Canada}

\begin{abstract}
	Cooperative collective dynamics is a principal determinant of the ability of synthetic micromotors to perform specific functions. However, realizing controllable and predictable collective behavior in complex physiological environments remains a significant challenge. Here, we show that collections of enzyme-coated colloids can be designed as various chemical logic gates, which subsequently can be organized into functional logic circuits. These circuits take environmental information as input signals and process it to produce output chemical species needed to achieve  specific goals. The chemical computation performed by the circuit endows the active colloidal system with the ability to sense its surroundings and autonomously coordinate its collective motion. The results of simulations of several examples are presented, where self-assembled colloidal circuits can identify invasive threats by their signals, produce and deliver chemicals to the targets to suppress their activity. The results of this work can aid in the design of experimental chemical logic circuits through micromotor self-assembly that autonomously respond to environmental cues to execute specific tasks.
\end{abstract}

\maketitle

\section*{Introduction}
\noindent
Synthetic micromotors form a distinctive class of active agents that have the potential to lead to transformative applications in the physical, chemical, and biological sciences~\cite{gompper2020roadmap,wang2013nanomachines,ju2025technology}. In biological applications, micromotors use self-propulsion to perform a number of functions such as actively targeted drug delivery or sensing and responding to invasive bodies ~\cite{lin2025hyperglycemia,hortelao2021swarming,wang2020self,tang2020enzyme,li2017micro,Reinisova2019micro}. Enzyme-powered motors feature prominently in many such applications because the chemical fuel they use is bio-compatible and bio-degradable, and experimental studies have shown how micro-colloids with Janus and random enzyme coats can be made using single enzyme species or multiple enzyme types~\cite{sattayasamitsathit2014dual,schattling2015enhanced,fernandez2020review,schattling2017double-fuel,ma2015enzyme,Patino2018influence,liu2023urease}.

Micromotors that sense their environments have been constructed; for example, laboratory studies have been carried out on enzyme-coated micromotors with pH-responsive nanovalves for on-demand cargo delivery~\cite{llopis2019enzyme} and with DNA nanoswitches for pH sensing~\cite{patino2019self}; also enzymatic cascades have been shown to produce chemomechanical communication~\cite{tseng2024chemomechanical}. Such investigations prompt the development of other active and autonomous sensing paradigms. In this paper, we show how logic-gate circuits constructed from colloid-based chemical logic
gates can be used to carry out computations that allow active colloidal systems to process and respond to information from their environment. The use of active colloidal chemical logic gates builds on the extensive literature dealing with the construction of chemical logic gates and logic-gate circuits in bulk phase systems~\cite{sugita1963circuit,arkin1994computational,seelig2006dna,niazov2006concatenated,unger2006towards,katz2010privmanrev,miyamoto2013synthesizing,chen2013programmable,katz2017review,katz2017enzymecircuits,gao2018programmable,chen2020novo,chen2021programmable,Winston2022gaterev,xu2025biocomputebook}. All standard logic gates can be built from protein enzymatic chemical reactions or from reaction kinetics that uses nucleic acid molecules. Logic-gate circuits may then be constructed through multiple coupled enzymatic logic gate reactions.
For example, various two-input gates using de novo designed proteins have been constructed~\cite{chen2020novo}, and have aided in programmable design of protein circuits ~\cite{chen2021programmable,chen2023Protein}.

The general principles used to design enzymatic gates in fluid phase systems can be applied to colloidal surface enzyme coats, enabling the construction of active-colloid-based logic circuits. Several features make the construction of colloidal gates built from protein enzymatic reactions feasible. Micron-scale colloidal particles can support protein coats on their surfaces comprising an order of $10^5$ molecules so that multiple surface enzymatic reactions, whose characteristics depend on the structure of the protein surface coat, can take place. Indeed, as discussed above, enzyme-coated active colloidal particles can support multiple enzymatic reactions. Models of such enzyme-coated colloids have been constructed, and simulations of their dynamics have shown that they can function as chemical logic gates~\cite{chen2024chemical}. Here, we describe how systems containing colloidal particles with different logic gates can act collectively to form logic-gate circuits that can be programmed to respond to a number of different chemical inputs to produce specific chemical outputs. Through this process, we show that the colloidal system can sense the chemical composition of its environment and autonomously respond to it to carry out a given task.

Although the operation of chemical logic-gate circuits in bulk phase systems with randomly dispersed enzymes is well documented, the localized nature of the enzyme coatings on active colloidal surfaces—and the consequent reshaping of signal transmission between individual colloidal logic units—necessitates investigation to determine how these logical circuits can form through self-assembly and encode environmental input to generate collective autonomous responses. The results presented in this paper address these issues and should serve as a guide for experiment.

The outline of the paper is as follows: Section~\ref{sec:circuits} begins with general considerations that underlie the construction of colloidal chemical logic-gate circuits. We then show how a colloidal logical circuit, with characteristics that are similar to an OR-AND-XOR circuit which was experimentally implemented in a microfluidic system~\cite{niazov2006concatenated}, can be made from three linked gated colloids, and summarize the results of simulations in the truth table that provides a Boolean representation of its operation. Through several examples in Sec.~\ref{sec:apps}, we show how versions of this and other circuits comprising unlinked colloids can cooperatively form by self-assembly and be used to target and neutralize invasive particles that emit chemical signals detected by the circuits. Further details on the construction of the colloidal circuits and the simulation methods used to study their dynamics are given in Methods and Supporting Information.

\section*{Results}
\subsection{Active colloidal logic-gate circuits}\label{sec:circuits}
The colloidal chemical logic gates that comprise a circuit are based on reaction kinetics that takes place on enzyme-coated colloid surfaces. Simple gates can be built on colloids with a uniform coat of a single enzyme species, or inhomogeneous colloid coats with enzyme-covered and inactive domains. More complicated gates that require several enzymatic reactions for their implementation can be built on colloids with inhomogeneous coats where domains have different enzymes. Such inhomogeneous coats can also be used to add a propulsive component to the gated colloids through self-generated chemical gradients that underlie phoretic mechanisms\cite{ruckner2007chemically}.

In order for the logic-gate colloids to form a circuit, the chemical output of one gate must be sensed as input for the next colloid gate in the circuit, enabling signal processing. Therefore, these colloidal collectives need to coordinate and cooperate with each other. In some situations, especially for biological systems where many reactions occur in the environment, chemical species are degraded by catabolic reactions as they diffuse, so the colloids should be in sufficiently close proximity to complete the circuit. This can be achieved in various ways. Diffusiophoretic motion of colloids in the chemical gradients that are generated by the action of the gates could lead to the self-assembly of the circuit. If the colloids respond to a localized source of a chemical species, diffusiophoretic motion of all the colloids in the circuit towards the source could lead to a cluster circuit that acts where it is needed. Linker molecules could be attached to the gates that selectively lead to the formation of a bound gated cluster that acts as a circuit. The mechanism that operates to form the circuit depends on the physical context.

\begin{figure}[!htbp]
	\centering
	\includegraphics[width=0.5\textwidth]{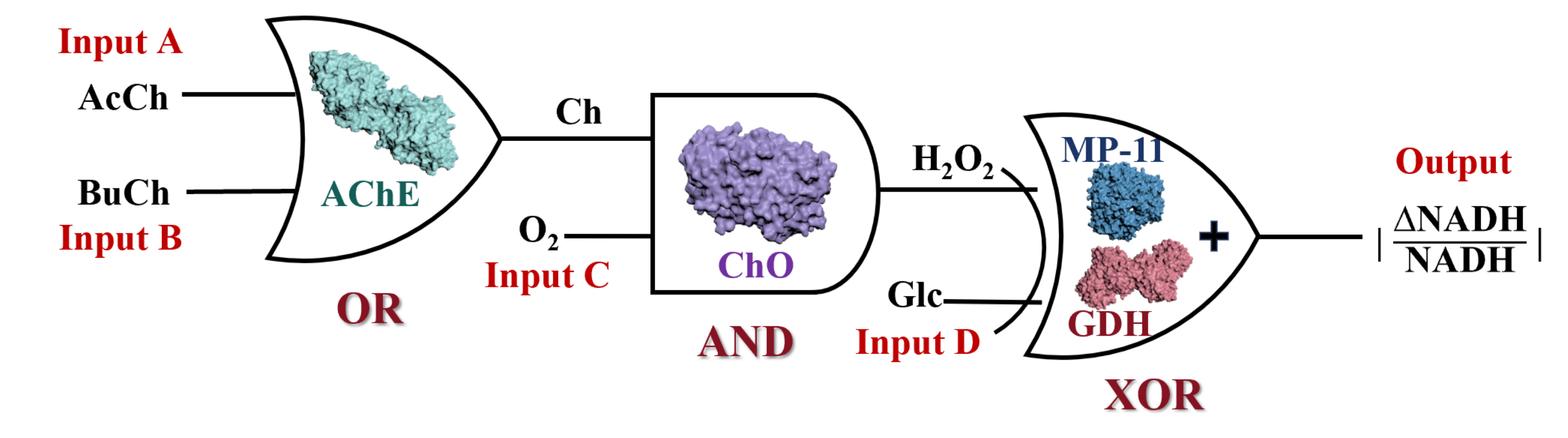}
	\caption{Schematic depiction of the operation of the OR-AND-XOR logic circuit constructed using four enzymes, AChE, ChO, MP-11, and GDH. Details of its operation are given in the text.}
	\label{fig:2-1}
\end{figure}

\vspace{0.25cm} \noindent
\textbf{\sf An OR-AND-XOR circuit}\\
To show how a colloidal logic-gate circuit can be made from various gates, we refer to experimental work on bulk phase enzymatic systems in which an OR-AND-XOR logic-gate circuit was constructed and studied~\cite{niazov2006concatenated}.
The experimental logical system makes use of four enzymes: acetylcholine esterase (AChE), choline oxidase (ChO), microperoxidase-11 (MP-11), and glucose dehydrogenase (GDH) coupled to its cofactor pair $NAD^+/NADH$. These enzymes catalyze the reactions,
\begin{eqnarray}
	\text{AcCh} &\xrightarrow{\text{AChE}}& \text{Ch} + \text{CH}_3 \text{COOH} \label{eq:or1}\\
	\text{BuCh} &\xrightarrow{\text{AChE}}& \text{Ch} + \text{CH}_3 \text{CH}_2\text{CH}_2 \text{COOH} \label{eq:or2}\\
	\text{Ch} + \text{O}_2 &\xrightarrow{\text{ChO}}& \text{Be} + \text{H}_2 \text{O}_2 \label{eq:and}\\
	\text{NADH} + \text{H}_2 \text{O}_2 &\xrightarrow{\text{MP-11}}& \text{NAD}^+ + \text{H}_2 \text{O} \label{eq:xor1} \\
	\text{NAD}^+ + \text{Glc} &\xrightarrow{\text{GDH}}& \text{NADH} + \text{GlcA},\label{eq:xor2}
\end{eqnarray}
and this enzymatic reaction network underlies the logic circuit. The circuit operates as follows (see Fig.~\ref{fig:2-1}): The OR gate is based on Eqs.~(\ref{eq:or1}) and (\ref{eq:or2}) where acetylcholine (input A) or butyrylcholine (input B) are hydrolyzed by AChE to form choline (Ch) that is the output of the first OR gate. (The ancillary products in these equations and elsewhere are not relevant for the operation of the circuit and their explicit dynamics need not be followed.) The AND gate based on Eq.~(\ref{eq:and}) processes the Ch product ($P_1$) produced by the OR gate, along with $O_2$ (input C) taken to be in excess, to yield betaine aldehyde (Be) and $H_2O_2$ as products ($P_2$). The two biocatalysts MP-11 and GDH are used in the more complicated exclusive OR (XOR) gate described by Eqs.~(\ref{eq:xor1}) and (\ref{eq:xor2}). The inputs for the XOR gate are $H_2O_2$ produced by the previous AND gate and glucose (Glc, input D). The operation of the circuit is read-out by following the concentration change ratio of the cofactor NADH ($F_2$), namely $|\Delta NADH(t)/NADH(0)|$ with $\Delta NADH(t)=NADH(t)-NADH(0)$. In the experiment, the output signal was monitored by the relative change in the absorbance of NADH and the results were used to construct the truth table for this logical circuit. Since the output is an analog signal, this circuit is made from fuzzy logic gates~\cite{zadegan2015fuzzydna,gentili2018fuzzymol,andreasson2015fuzzy}. The gated action of enzymatic reactions is determined by the forms of the enzyme response curves~\cite{arkin1994computational} or by filtering in experiments~\cite{zavalov2012or-sigmoid}. In addition, conditions on the number of enzymatic gates that can be used effectively in logic circuits have been determined~\cite{privman2008many-gates}.

\vspace{0.25cm} \noindent
\textbf{\sf Construction of colloidal logic circuits and simulation of dynamics}\\
The colloidal logic circuit we implement in our simulations is constructed from a simpler reaction network that focuses on the main species involved in the operation of the gates:
\begin{eqnarray}
	OR:&& \quad \left\{\begin{array}{c}
		A \xrightarrow{E_1} P_1 \\
		B \xrightarrow{E_1} P_1
	\end{array} \right., \label{eq:OR}\\
	AND:&& \quad \quad  C^*+ P_1 \xrightarrow{E_2}  P_2, \label{eq:AND} \\
	XOR:&& \quad \left\{\begin{array}{c}
		P_2 + F_2 \xrightarrow{E_3} F_1 + S  \\
		D + F_1 \xrightarrow{E_4} F_2 + S \label{eq:XOR}
	\end{array} \right. .
\end{eqnarray}
The input species $\{A,B,C,D\}$ and the intermediate chemical species $P_1$ and $P_2$ are processed by this network in a way that is similar to that of the inputs and the $\text{Ch}$ and $\text{H}_2 \text{O}_2$ species in the experimental network. The concentration of C, like $\text{O}_2$ in the experiment, is assumed to be constant and its value is incorporated in the reaction rate coefficient for enzyme $\text{E}_2$. (Here and elsewhere, species whose concentrations are in excess are indicated by the superscript * as in $C^*$.) This input is assigned the Boolean  value of 1.  In addition, to implement the effects of open-system kinetics, we make use of a species $S$ that undergoes reactions $S \xrightarrow{k_{cat}} A/B/D$ in the fluid phase. These reactions mimic the effects of reservoirs that supply chemical species to the system to maintain it in a nonequilibrium state, and allow the circuit to operate in the steady state regime since the input species are uniformly regenerated in the environment to compensate for their consumption in the network reactions.

\begin{figure}[!htbp]
	\centering
	\includegraphics[width=0.5\textwidth]{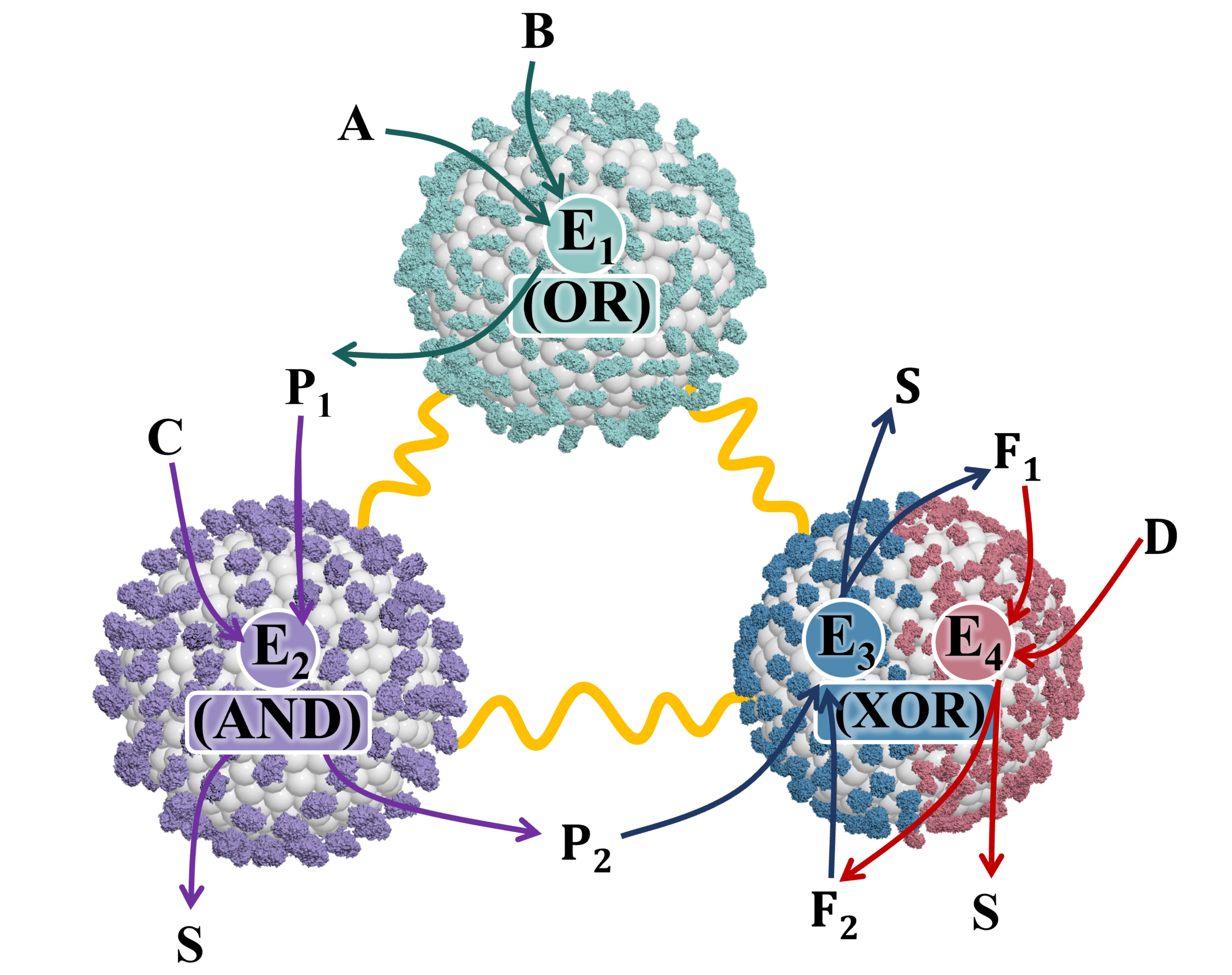}
	\caption{The OR-AND-XOR colloidal logic-gate circuit. The circuit uses three linked colloids with coats constructed from four different enzymes as described in the text. The figure shows the reactions that underlie the OR, AND, and XOR colloidal gates. The gated colloids reside in a reactive chemical medium.}
	\label{fig:2-2}
\end{figure}

For this example, we propose a colloidal circuit formed from linked colloids, and the configuration of the three colloids used in our simulations is shown in Fig.~\ref{fig:2-2}. To model the linker groups, the centers of mass of the colloids are connected by stiff harmonic springs to keep them in proximity so that the circuit can function effectively. Uniform coats of enzymes $E_1$ and $E_2$ are on two of the colloids. On the surface of the third colloidal sphere, one hemisphere is coated with enzyme  $E_3$, while the other hemisphere is coated with enzyme $E_4$. The colloid complex, along with the fluid particles, reside in a simulation box with periodic boundary conditions. The system evolves by hybrid molecular dynamics-multiparticle collision dynamics~\cite{malevanets1999mpc,malevanets2000mpchybrid,kapral2008multiparticle,gompper2009multi} as described in Methods and Supporting Information where full simulation details are provided.

Because the XOR gate is constructed from a colloid with a Janus coat of two different enzymes on the two hemispheres, this colloid could serve as a propulsive component for the linked cluster configuration. We could have constructed this gate from a random distribution of the two enzymes on the surface of the colloid. In fact, depending on the interaction potentials of the reactive species with the enzymes, even if all colloids had uniform enzyme coats, active motion of the complex is possible through a mechanism similar to that of sphere dimer motors where one sphere responds to concentration gradients produced by its linked partner~\cite{ruckner2007chemically}. Here, we have chosen the fluid-colloid interaction potentials to be the same for all species, so there is no active motion. This allows us to focus on how the circuit processes chemical signals, our primary objective in this section. In the applications presented in the next section, the potential energy interactions will not be identical for all species, and harmonic constraints will not be placed on the colloids so that they will be free to move and respond to chemical gradients to form cluster circuits.

\begin{figure}[!htbp]
	\centering
	\includegraphics[width=0.4\textwidth]{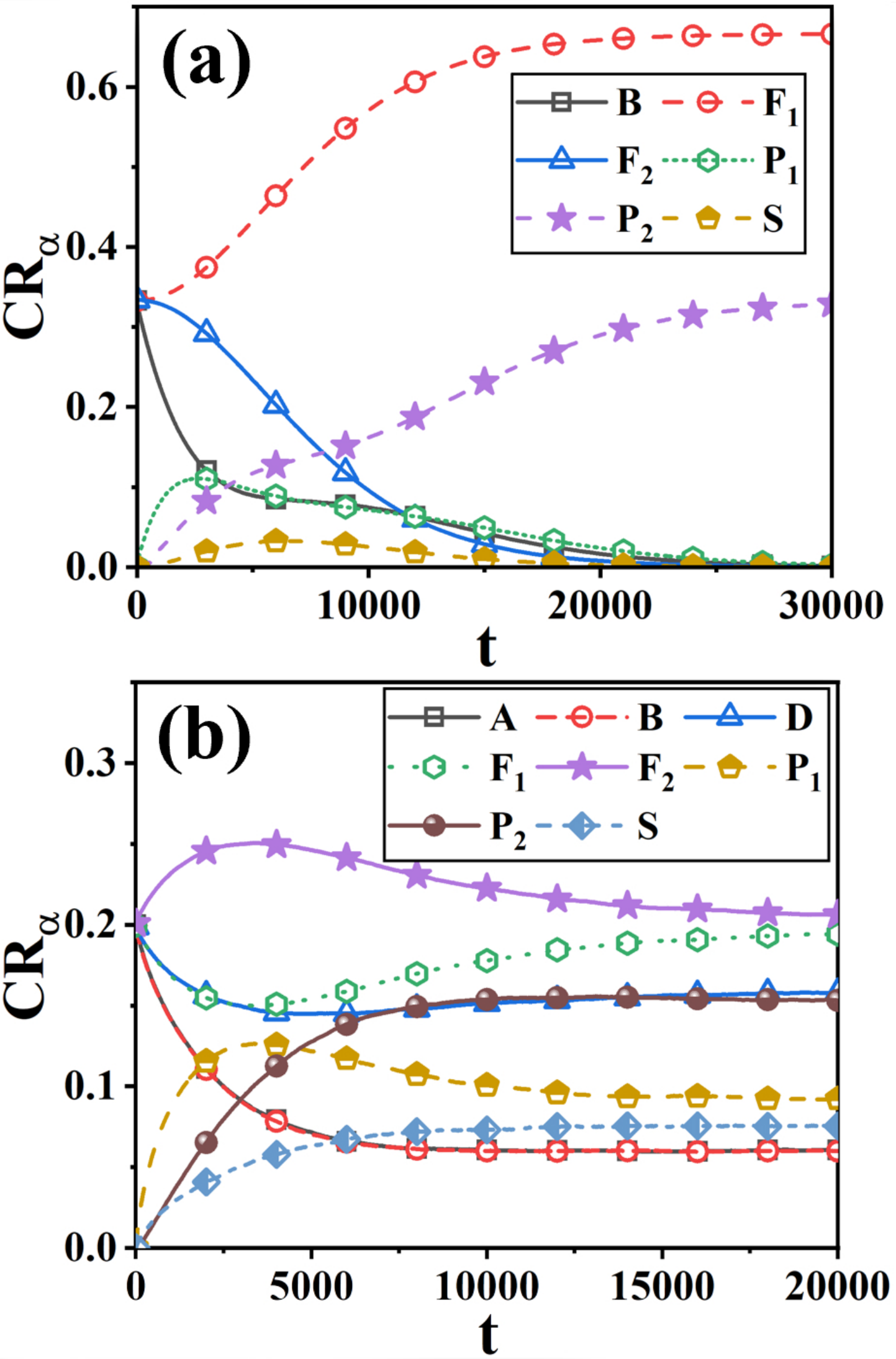}
	\caption{Temporal evolution of species concentration ratios $CR_\alpha$ in the system for two (ABCD) inputs. (a) Input (0110) results in output signal ``1''. (b) Input (1111) results in output signal ``0''. The concentration ratio $CR_\alpha = N_\alpha(t)/N_0$, where $N_\alpha(t)$ is the number of particles of species $\alpha$ at time $t$ ($\alpha = A, B, D, F_1, F_2, P_1, P_2, S$) and $N_0$ is the total particle number in the system.}
	\label{fig:4}
\end{figure}

The output of the OR–AND–XOR logic circuit to various input signals is taken to be the  ratio of the concentration change of $F_2$ to its initial value, $\Phi(t)=|\Delta F_2(t)/F_2(0)|$. A value of $\Phi(t)$ less than 0.5 is assigned a Boolean value  of “0”, while higher values are assigned “1”, so, as noted earlier, the gates we consider are fuzzy logic gates. The truth table for the circuit can be constructed from simulations of the dynamics that yield outputs for the eight possible inputs to the circuit. As examples, the temporal evolutions of the concentrations of all active species are shown in Fig.~\ref{fig:4} for two input values to the circuit: (0110) and (1111).

\begin{figure}[!htbp]
	\centering	
	\includegraphics[width=0.45\textwidth]{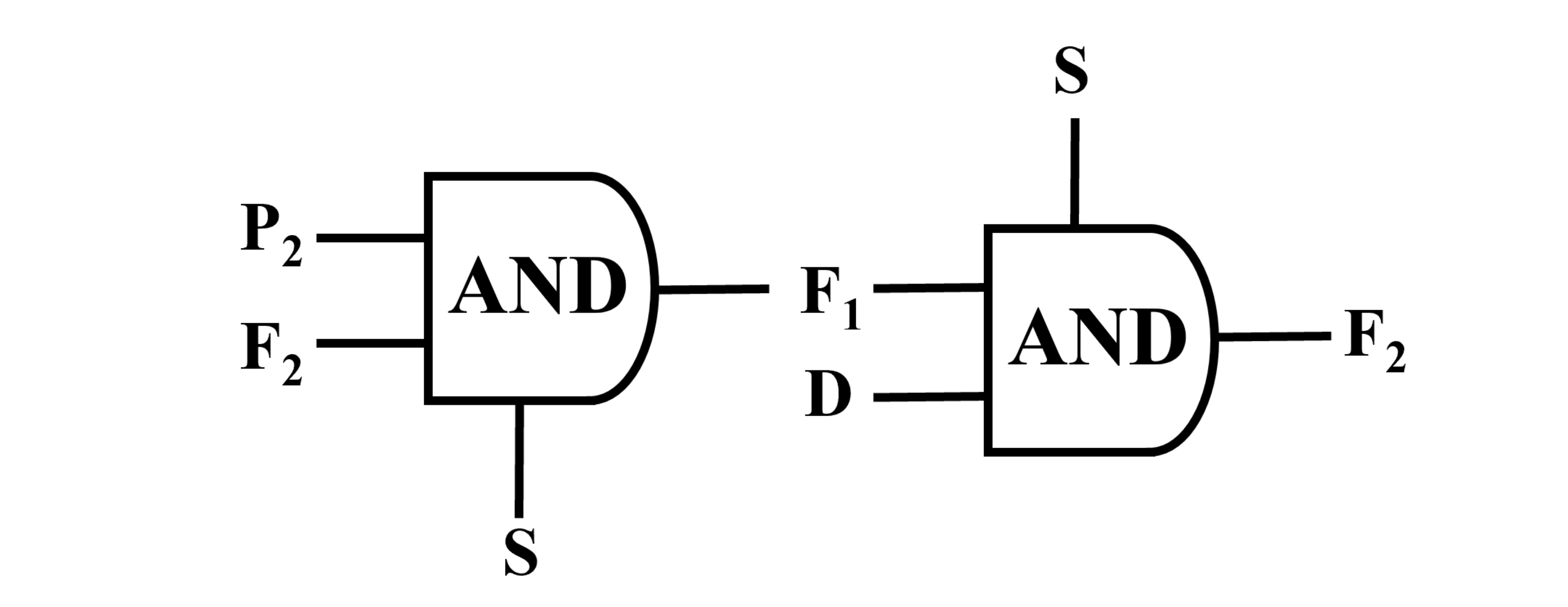}  
	\caption{The XOR gate constructed from two AND gates.}
	\label{fig:XORastwoAND}
\end{figure}

Since the input C is always present and has value 1, there are eight possible inputs to the circuit specified by $(AB1D)$. Consider the input (0110) where the B species is the only other input to the circuit. Recall that for this model species $F_1$ and $F_2$  are also present in the solution. Species B activates the OR gate through the surface catalytic activity of enzyme $E_1$, and this leads to the generation of product $P_1$ via reactions described by Eq.~(\ref{eq:OR}). As observed in Fig.~\ref{fig:4} (a), the concentration of B decreases, while that of $P_1$ exhibits a transient rise. The second logic gate in the circuit coated by $E_2$ acting as an AND gate converts $P_1$ to $P_2$ through the reaction described by Eq.~(\ref{eq:AND}). After reaching a maximum, the concentration of $P_1$ decreases, while $P_2$ continues to increase. Species $P_2$ and $F_2$ initiate the reaction described by Eq.~(\ref{eq:XOR}), activating the XOR gate: enzyme $E_3$ on the Janus surface converts $F_2$ to $F_1$. We note that the XOR gate can be constructed from the combination of two AND gates as shown in Fig.~\ref{fig:XORastwoAND}, and only the first of the two AND gates is activated with this input since species D is absent. The fluid phase reaction $S \xrightarrow{k_{cat}} B$ continuously supplies B until all $F_2$ is converted to $F_1$. The Boolean output corresponding to input (0110) is 1.

\begin{figure*}[t]
	\centering	
	\includegraphics[width=\textwidth]{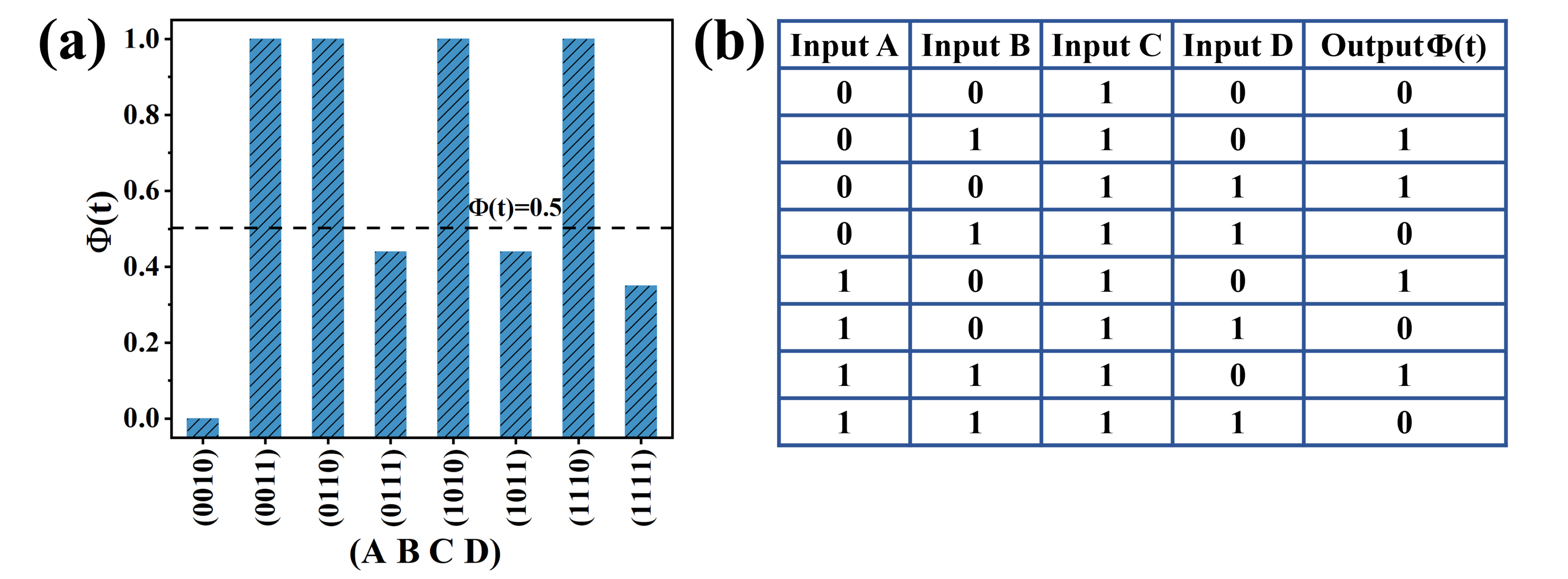}  
	\caption{(a) Bar presentation of the outputs  of the OR-AND-XOR circuit. Each bar corresponds to a specific input and its corresponding output signal, $\Phi(t)$. The horizontal dashed line at $\Phi(t)=0.5$ denotes the threshold between high and low $\Phi$, specifying the values of the output signals as ``0" and ``1" respectively. (b) Truth table for the the circuit. }
	\label{fig:truth-table}
\end{figure*}
When the input signal is (1111), species A, B and D are present, and all enzymes on the colloid surfaces are involved in the operation of the logical circuit. The evolution of the concentrations of all species is shown in Fig.~\ref{fig:4} (b). The concentrations of A, B and D initially decrease, then gradually approach a steady state due to the bulk phase reaction $S \xrightarrow{k_{cat}} A/B/D$. The concentration of the intermediate product $P_1$ increases rapidly, since it is generated from the conversion of A and B in the OR gate reaction, and then decreases to a steady state value due to consumption at the $E_2$ AND gate. Species $P_2$ is generated and consumed in reactions (\ref{eq:AND}) and (\ref{eq:XOR}), and its value increases to a final steady state. Species $F_1$ and $F_2$ reach a steady state that is approximately the same as the initial state, so the output of this state from $\Phi(t)$ has Boolean value 0.

The output signals from the remaining six inputs are shown in the Supporting Information. The results are summarized in Fig.~\ref{fig:truth-table} (a), and the corresponding truth table is presented in Fig.~\ref{fig:truth-table} (b). Although the dynamics underlying the OR-AND-XOR colloidal circuit differs from that in the bulk-phase experiment~\cite{niazov2006concatenated}, the truth table has the same form. Therefore, colloid-based enzyme logic circuits can effectively process environmental signals.

\begin{figure*}[b]
	\centering	
	\includegraphics[width=\textwidth]{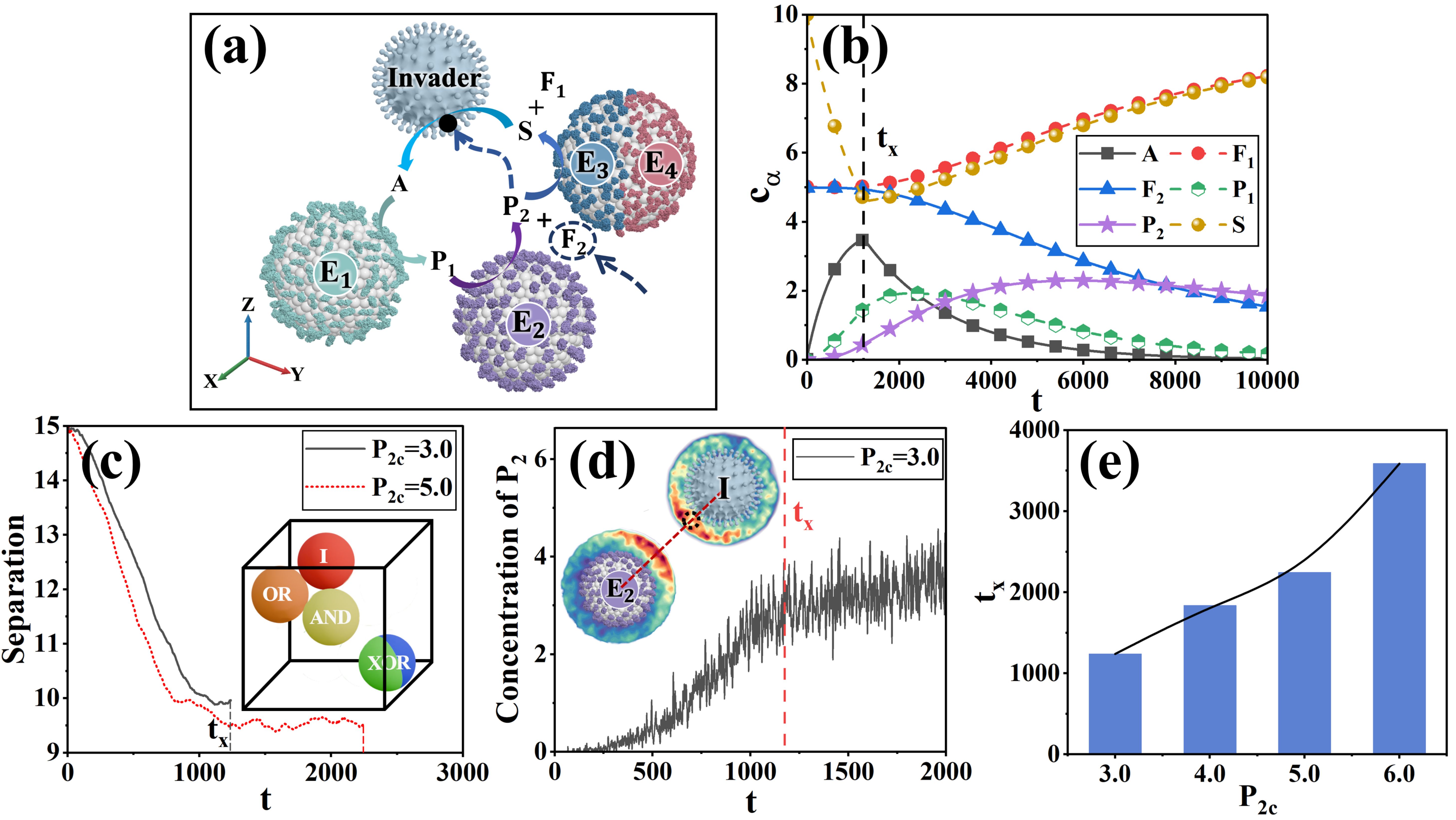}  
	\caption{(a) Schematic diagram shows the kinetics of the enzymatic logic-gate circuit responding to an Invader. The $S \to A$ chemical activity of $I$ can be inhibited by a species $P_2$ when its concentration exceeds a critical value, as indicated by the small solid black circle in the diagram. (b) Temporal evolution of the six species concentrations $c_{\alpha}$ in the system. The vertical dashed line marks the lifetime $t_{\rm x}$ of $I$ when its activity is terminated. (c) The evolution of the separation between the AND colloid ($E_2$) and $I$ at different critical concentrations $P_{2c}$.  The inset shows the positions of all colloids in the system at the moment $I$ loses activity. (d) The temporal evolution of the concentration $c_{P_2}$ of $P_2$ around $I$ (plotted from the dashed circular area near $I$ in the inset). The inset shows the concentration field of $P_2$ around $I$ ($4.0<r<7.0$), and of A around $E_2$ ($4.0<r<7.0$). The symbol $P_{2c}$ denotes the critical concentration of $P_2$ that causes $I$ to lose its activity. (e) The lifetime $t_{\rm x}$ for several values of the critical concentration $P_{2c}$. }
	\label{fig:Invader-1}
\end{figure*}

\subsection{Detection and suppression of invader activity}\label{sec:apps}
Often a goal in biomedical applications is selective targeting of an invasive body in order to terminate its activity \cite{Han2026Intratumoural}. Targeting diseased cancer cells by the different biomarkers they produce is an example of such an application, and for this purpose, OR, AND and NOT protein logic gate switches have been used to distinguish cells that express different specific antigens.~\cite{lajoie2020target-cells} Different types of invasive agents may give rise to distinct environmental chemical signals that are recognized by active colloidal systems, thereby forming logic circuits to respond to the threats. In this section, we provide several examples that show how colloidal logic-gate circuits can be used to perform similar tasks. In particular, we consider situations where invasive particles that are harmful to the system are present and demonstrate that self-assembled colloidal logic-gate circuits can be used to remove the threats posed by these invaders.

\vspace{0.25cm} \noindent
\textbf{\sf Single invasive species}\\
{\bf Example 1:} We suppose that the system contains a single invasive particle (termed as Invader $I$) that consumes chemical fuel S in its environment and produces a harmful product A,S which also signals the presence of the invader in the system. The activity of the Invader can be inhibited by a chemical $P_2$ that interferes with the reaction that sustains it by using an inhibitor (INH) gate. Species $P_2$ is not present in the environment; however, the system does contain a number of (unlinked) gated colloidal particles that support different logic gates. Among these are the OR, AND and XOR gates discussed above; however, we regard the gate reactions in Eqs.~(\ref{eq:OR})- (\ref{eq:XOR}) as generic reactions for such gates without specific reference to the experimental circuit that prompted the model equations. The Invader and the colloidal gates involved in this application are shown in Fig. ~\ref{fig:Invader-1} (a).

\begin{figure}[!htbp]
	\centering	
	\includegraphics[width=0.45\textwidth]{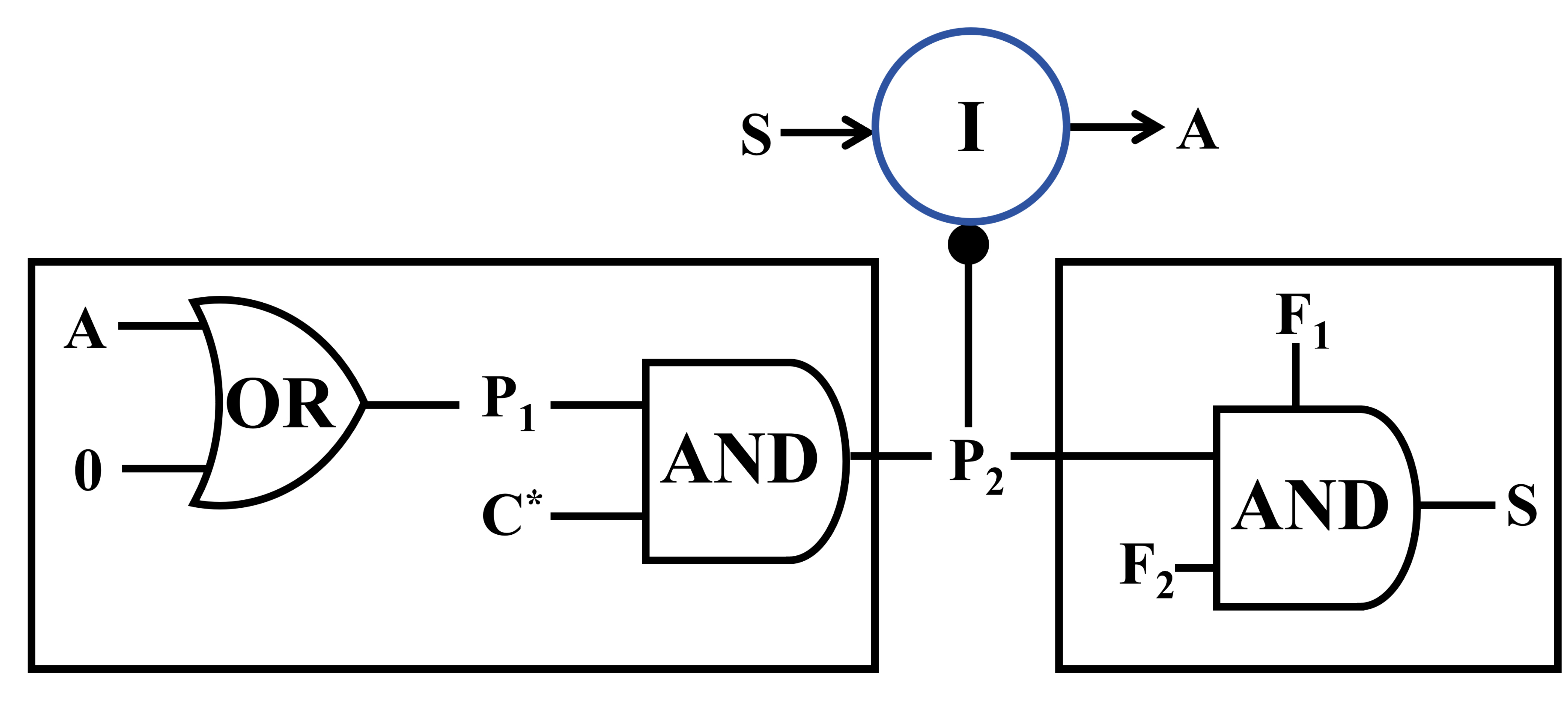}  
	\caption{The OR-AND-AND-INH circuit used in this model. }
	\label{fig:or-and-and-ihn}
	\vspace{-5mm}
\end{figure}

Initially, the system contains only three types of particles in solution: $S$, $F_1$, and $F_2$, the latter two involved in the operation of the XOR gate. Recall that the XOR gate can be constructed from two AND gates and, since species D is not present, only the AND gate involving enzyme $E_3$ is active. The activated circuit used in this application is shown in
Fig.~\ref{fig:or-and-and-ihn}.

The colloidal system can detect the presence of the Invader by its chemical signal A, and this signal triggers a response that leads to the formation of a circuit that suppresses the activity of the Invader. From Fig.~\ref {fig:or-and-and-ihn} one can see that the OR-AND circuit in the left box will generate the $P_2$ species that is needed. The evolution of the species concentrations are as follows (Fig. ~\ref{fig:Invader-1} (b)): The initial  increase in the concentration of A and a concomitant decrease in S due to the presence of the Invader produces a concentration gradient of A around $I$. Once the OR colloid senses the  A gradient, it experiences positive chemotactic motion toward $I$, since A is a substrate for $E_1$. In the simulation, this directed migration is realized by a diffusiophoretic mechanism involving specific energy parameters (see Supporting Information). The activated logic circuit generates $P_1$ on the OR gate. Likewise, the AND colloid senses the gradient field of $P_1$ around the OR colloid and moves closer toward it, and they approach $I$ together. Fig.~\ref{fig:Invader-1} (c) shows the decrease in the separation between $I$ and the AND colloid. The organization of the OR-AND part of the circuit occurs through substrate-driven chemotaxis~\cite{sapre2025non,somasundar2019positive}. In such cascade chemotaxis each enzyme independently follows its own substrate gradient, which in turn is produced by the preceding enzymatic reaction in the circuit~\cite{zhao2018substrate}. As shown in Fig.~\ref{fig:Invader-1} (d), this process results in an increase in the local concentration $c_{P_2}$ of $P_2$ in the vicinity of $I$, and when the concentration of $P_2$ exceeds a critical threshold $P_{2c}$, the activity of $I$ is suppressed by the INH gate.  If the value $P_{2c}$ required for suppression $I$ is increased, the AND colloid must move closer to $I$, resulting in a smaller separation, as seen in Fig.~\ref{fig:Invader-1} (c). Similarly, the active lifetime $t_{\rm x}$, from the entry of $I$ into the system until its chemical activity is suppressed, also increases,as shown in Fig.~\ref{fig:Invader-1} (e).

Although the task of suppressing the activity of the invader has been accomplished, the system is still left with an unused concentration of $P_2$, an unwanted possibly harmful agent such as $H_2O_2$. The second AND colloidal gate in the right box in Fig.~\ref{fig:or-and-and-ihn} is then used to remove this species. The continuous production of S through this pathway effectively replenishes the originally-present S molecules in the bulk solution. In addition, this newly generated S establishes a feedback loop: it diffuses towards the Invader and serves as the precursor for the sustained generation of A, thereby maintaining the stable A and $P_1$ concentration gradients that are essential for OR ($E_1$) and AND ($E_2$) sensing, respectively.

\begin{figure*}[t]
	\centering	
	\includegraphics[width=\textwidth]{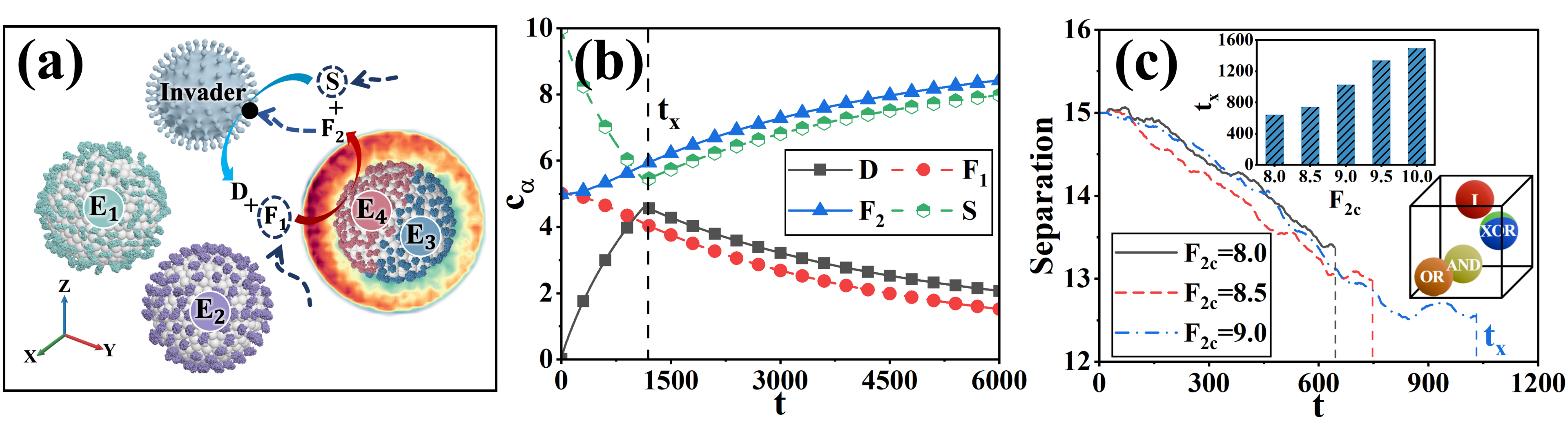}  
	\caption{(a) Picture of the logical circuit activating a portion of the XOR colloid in response to an invasive species that emits species D. The $S \to D$ activity of $I$ is inhibited by the action of an INH gate for a high ratio of $F_2/F_1$ concentrations in the domain ($4.0<r<7.0$) around the Invader. The asymmetric concentration profile of $D$ around the XOR Janus colloid is also shown in the figure.  (b) Time-dependent evolution of four species concentrations in the system. (c) The evolution of the separation between the Janus colloid and $I$ at different critical concentrations $F_{2c}$. The upper middle inset shows the lifetime $t_{\rm x}$ as a function of $F_{2c}$, while the lower right inset shows the colloidal configuration.}
	\label{fig:invader-2}

\end{figure*}
\begin{figure*}[t]
	\centering	
	\includegraphics[width=\textwidth]{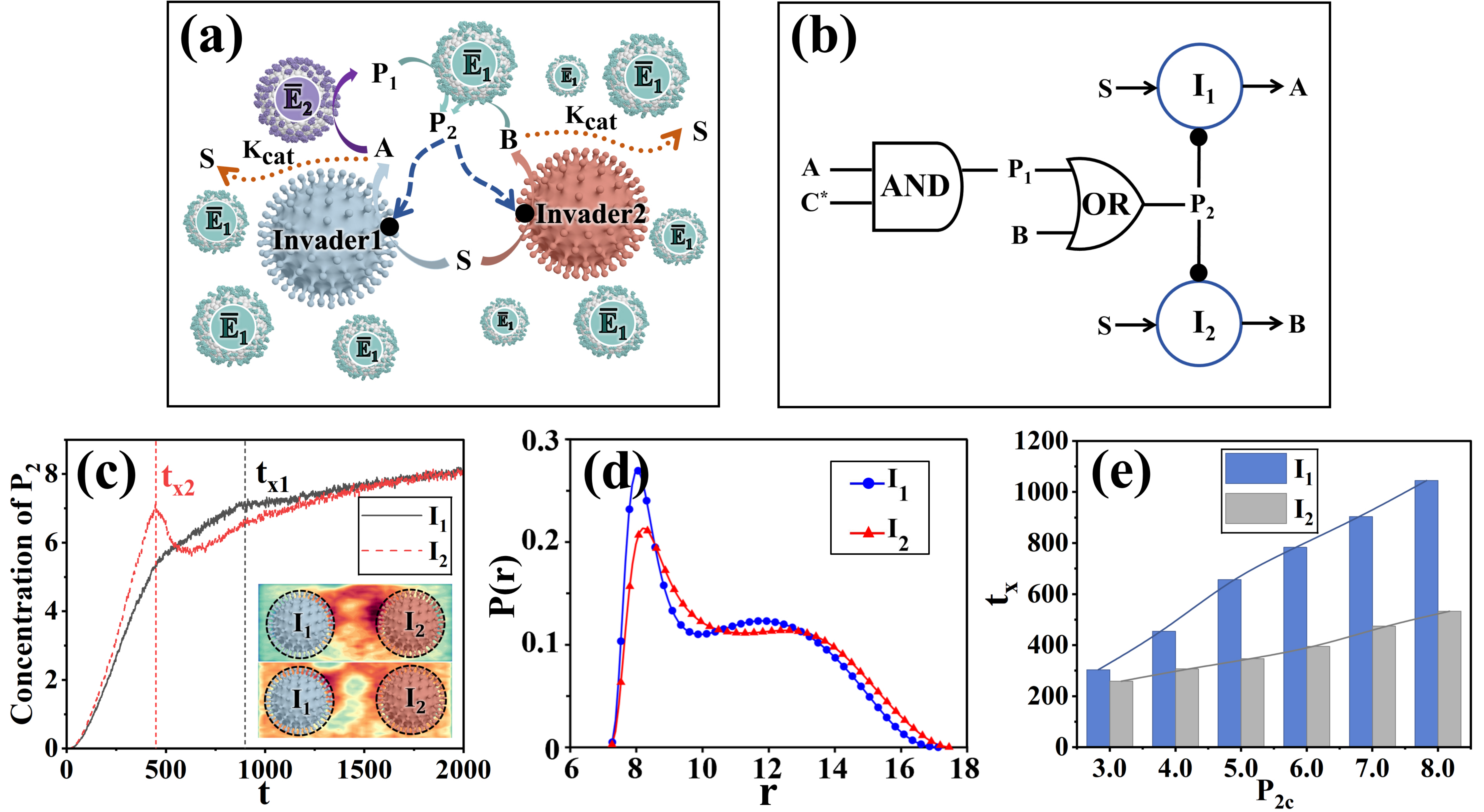}  
	\caption{(a) Diagram shows the kinetics of the enzymatic logic-gate circuit responding to two different Invaders, $I_1$ and $I_2$.  The simulation system contains one AND colloid coated by $\overline{E}_2$ and eight OR colloids coated by $\overline{E}_1$. (b) The AND-OR-INH circuit used in this model. (c) The temporal evolution of the concentration $c_{P_2}$ around $I_1$ (solid lines) and $I_2$ (dashed lines). These concentrations are averages over spherical shells ($4.0<r<4.5$) surrounding $I_1$ and $I_2$, respectively. The insets show the concentration profiles of $P_2$ around $I_1$ and $I_2$ at t=$t_{\rm x2}$ (upper) and t=$t_{\rm x1}$ (lower), respectively. (d) The radial probability distribution $P(r)$ of OR-gates around $I_1$ and $I_2$ at t=500. The results are determined from 20 realizations of the dynamics. (e) The dependence of the lifetimes $t_{\rm x1}$ and $t_{\rm x2}$ on the the critical concentration $c_{P_{\rm 2c}}$.}
	\label{fig:two-invaders}
\end{figure*}

{\bf Example 2:} In the previous example, the logical circuit was assembled in response to an input of A (input signal (1010); recall C takes Boolean value 1). Other inputs to a system containing colloids with he same gates can elicit different responses. For example, suppose that the fluid phase has nonzero initial concentrations of S, $F_1$ and $F_2$  species, with $c_{F_1}=c_{F_2}$. At $t=0$ an invasive species that consumes S and produces a signal chemical D is introduced into the system. The activity of the invader will be suppressed if the concentration ratio $F_2/F_1$ is sufficiently high. The gated colloidal system will respond in a simple way to such a threat. The input signal for species D  (0011) will activate the right AND gate in the composite XOR gate shown in Fig.~\ref{fig:XORastwoAND}, with other gates remaining inactive. The system is shown in Fig.~\ref{fig:invader-2} (a).  As time evolves, the concentration of $F_2$ increases (Fig.~\ref{fig:invader-2} (b)), thereby increasing the $F_2/F_1$ ratio.

Concentration gradients of D are established around $I$ and the XOR Janus colloid. These gradients cause the XOR colloid to move closer to $I$ through substrate-driven chemotaxis. Fig.~\ref{fig:invader-2} (c) shows how the separation between the Janus XOR colloid and $I$ decreases. The gate reaction on $E_4$ results in a higher concentration of $F_2$ surrounding $I$ (Fig.~\ref{fig:invader-2} (a). When the $F_2/F_1$ ratio exceeds the critical concentration $c_{F_{2c}}$, the catalytic activity of $I$ is inhibited by an INH gate as in the previous example. The inset on the lower right in Fig.~\ref{fig:invader-2} (c) shows an instantaneous configuration of all colloidal particles, where the close proximity of the Janus colloid to the $I$ particle can be seen. The upper middle inset of this figure shows how $t_{\rm x}$ increases with increasing $F_{2c}$ concentration.

In the Supporting Information we give other examples, including cases where the invasive body produces more than a single signal chemical species.

\vspace{0.25cm} \noindent
\textbf{\sf Two invasive species} \\

In some circumstances, the system may be exposed to more than a single type of invasive species. In this section, we show that ensembles of enzyme-coated colloidal motors can be configured to implement distinct logic circuits that collectively respond to and suppress multiple invaders. In particular, we consider situations where two different invasive species are introduced into the system: Invader 1 ($I_1$) and Invader 2 ($I_2$) can convert S-particle fuel in the environment to signal A and B particles, respectively. An enzyme-based logic circuit suppresses the activity of both invader species. The reactive fluid system contains a single colloid functionalized with enzyme 2 ($\overline{E}_2$) that acts as an AND gate and eight colloids functionalized with enzyme 1 ($\overline{E}_1$) that act as OR gates, together with invader species. The overlines on enzymes are introduced to stress that enzymes used to build AND and OR gates are not the same as those for the single invader examples.

The colloidal reaction network is shown in Fig.~\ref{fig:two-invaders} (a) and the corresponding simplified logic circuit is presented in Fig.~\ref{fig:two-invaders} (b): when particle A encounters the AND gate with $\overline{E}_2$, it is converted to $P_1$ in the presence of species C; subsequently $P_1$ is transformed into $P_2$ on the OR gates with $\overline{E}_1$. The OR gate colloids can also convert particle B into $P_2$. The product $P_2$ then inhibits the activity of the invaders through the action of an INH gate. Particles A and B are degraded to S in the environment by catabolic reactions $A/B \xrightarrow{k_{cat}} S$ on long timescales.

The colloidal circuit is completed by self-assembly through substrate-driven chemotaxis, similar to that for a single invasive species: the AND gate co-migrates with the downstream OR-gate colloids toward $I_1$, resulting in a rapid increase in the local concentration $c_{P_2}$ of $P_2$ around $I_1$ (solid lines in Fig.~\ref{fig:two-invaders} (c)). Since B also functions as the direct substrate for the OR-gate colloids, they also move by chemotaxis toward $I_2$, thus increasing $P_2$ concentration in the vicinity of $I_2$ (dashed lines in Fig.~\ref{fig:two-invaders} (c)). Consequently, the OR-gate colloids can accumulate either around $I_1$ or $I_2$ depending on local environmental conditions.

Compared to the increase of $P_2$ around $I_1$ induced by AND~($\overline{E}_2)-OR~(\overline{E}_1)$ cascade chemotaxis, the direct chemotactic response of OR-gate colloids to their other substrate B leads to a more rapid accumulation of $P_2$ around $I_2$. This is evident in the $c_{P_2}$ curves in Fig.~\ref{fig:two-invaders} (c) where the concentration of $P_2$ is higher around $I_2$ than $I_1$. When identical critical thresholds $P_{2c}$ are applied to both $I_1$ and $I_2$, for the given simulation parameters, $I_2$ is inhibited earlier than $I_1$ ($t_{\rm x2}<t_{\rm x1}$). When the chemical activity of $I_2$ is suppressed at $t=t_{\rm x2}$, the production of B stops, leading to a rapid decrease in $P_2$ concentration around $I_2$ (dashed-line curve in Fig.~\ref{fig:two-invaders} (c)).  The $P_2$ concentration around $I_1$ shows the opposite trend (cf. concentrations at $t=t_{\rm x2}$ and $t_{\rm x1}$ (lower) in the insets). The OR-colloid probability density shown in Fig.~\ref{fig:two-invaders} (d) is bimodal, indicating accumulation of OR colloids near both $I_1$ and $I_2$. Consequently, the OR colloids in the logic circuit self-organize as a result of changes due the invasive species. Moreover, Fig.~\ref{fig:two-invaders} (e) shows that increasing the value of $P_{2c}$ prolongs the lifetimes of both $I_1$ and $I_2$, while the lifetime gap, $t_{x1}$-$t_{x2}$, also widens.

\section*{Discussion} \label{sec:conc}
The research reported in this paper showed how logical circuits constructed from colloid-based chemical logic gates can be used to sense chemical signals and respond to those signals by performing specified functions. Chemical logic gates can be built using enzyme or nucleaic acid reaction kinetics, as demonstrated in previous experimental investigations of bulk-phase and micro-fluidic systems. Enzymatic reactions are especially convenient for this purpose because active colloids with various enzyme coats have been made and their properties have been studied.

Colloidal logic circuits are built by coupling chemical reactions on different gated colloids to form a chemical network that can process chemical signals. In order for the circuit to form and function, the colloids that comprise the circuit should be in close proximity, so the output of one gate could serve as input to another gate. This can be achieved by design through specific linking molecules, substrate-driven diffusiophoretic motion or active self-propulsion. The operation of the fuzzy logical circuits formed in this way can be classified in terms of the Boolean operations they perform, so that the output of the circuit is the result of a chemical computation.

The dynamical simulations have served to highlight two aspects that determine how colloidal chemical logic-gate circuits operate: processing chemical signals, and circuit assembly and dynamics related to function. Once a circuit is specified by its constituent gates and their connections, its output for given inputs is encoded in the truth table (cf. Sec.~\ref{sec:circuits}). For circuits made of colloidal gates, this requires the selection of coated colloids that catalyze specific chemical reactions that underlie the different gate functions. The gate reactions in the circuits then form a chemical network that processes input chemical signals to yield the desired output species. The logic gate structure that underlies the chemical network and the Boolean representation of the chemical inputs and outputs provides a way to construct the chemical networks needed in applications~\cite{solanki2023functions}. Thus, the representation of the chemical network in terms of logical circuits comprising chemical logic gates provides an organizing structure for the  often-complex chemical kinetics that these active systems possess.

Considering general scenarios that utilize colloidal logic gates, instead of initially linking the colloids to form a circuit, the system should be able to respond to input by autonomously self-assembling the colloids needed to build the circuit. For this purpose, one must make use of the intermediate chemicals involved in the circuit and its active properties, rather than only its final chemical output. Substrate-driven chemotaxis can play an important role in such self-assembly, but so can colloid self-propulsion and dynamical linking. Once the circuit is constructed and operational, the chemical output must be put to use. This could involve movement of the colloidal circuit to the vicinity of some target, as was the case for the destruction of invasive species discussed in Sec.~\ref{sec:apps}. Although we described how self-propulsion of gated colloids can be included in the dynamical description of colloidal circuits, our examples did not make use of this effect. Other applications, such as targeted pick-up and delivery of cargo, could use different strategies involving the self-propulsion of the colloids comprising the circuit.

It should be possible to construct chemical logic-gate systems whose response is not limited to a specific application. For instance, consider a system of many active gated colloidal particles supporting a number of different logical gates. Such a system can process many different input signals and, when presented with  a specific chemical signal, it will respond by assembling a prescribed subset of gated colloids to yield a corresponding output. The nature of the collective dynamics of such large ensembles of gated colloids and their responses to input chemical signals merit further study.

\section*{Methods}
This Methods section provides a qualitative summary of relevant aspects of the simulation methods used to obtain the results in the main text. It also serves as a guide to the complete description of all simulation details that are given in the Supporting Information (SI).

\textbf{Colloid model}
A spherical colloid is constructed from point particles that are randomly distributed inside a sphere with radius $R$, along with coarse-grained enzyme beads that are uniformly distributed on the spherical surfaces of the colloid. If the colloid has an enzyme coat comprising two more enzymes or inactive surface beads, non-uniform distributions that depend on the physical context are implemented. The point particles and the enzyme beads are connected by  harmonic springs. Invader species are modeled by single spheres with radius $R_I$.

\textbf{Enzyme logic gate models}
Enzyme-catalyzed reactions on the surface of a colloid are carried out using a particle-based model. When a solution substrate reactive particle enters the reaction zone of an enzyme bead, a product particle is formed with a specified probability. The reactive dynamics conserves mass, momentum and energy. The reaction rule depends on the identity of the logic gate; for instance, an AND gate has two inputs, and the reaction rule is constructed so that the product is produced only when two substrates are in the reaction zone. In addition, since the reactions are gated, restrictions are placed on the local concentrations of substrate species to enforce gate conditions.

\textbf{System dynamics}
The dynamics of the entire system is described by hybrid Molecular Dynamics-Multi-Particle Collision (MD-MPC) dynamics~\cite{malevanets1999mpc,malevanets2000mpchybrid,kapral2008multiparticle,gompper2009multi}. In MPC dynamics interactions among the fluid particles are taken into account through multparticle collisions in cells with linear dimension $a_0$. In the MD steps the system evolves by Newton’s equations of motion with forces determined from potential functions. Enzymatic catalytic reactions  occur at each MD step. Fluid phase reactions are described by reactive multiparticle collision dynamics~\cite{rohlf2008reactivempc}. The simulation volume is a three-dimensional box with volume $V = L_x \times L_y \times L_z$, which contains the colloids along with a large number $N_S \sim 10^6$ of explicit solution molecules.

\textbf{Interaction potentials}
The interaction potential between solvent particles and
colloid surface enzyme beads is either a repulsive Lennard-Jones (LJ) potential, or a truncated attractive LJ potential, depending on the application. The colloids interact with each other through repulsive LJ potentials, unless linked colloid configurations are used. These interactions are characterized by energy $\epsilon$ and distance $\sigma$ parameters.

\textbf{Dimensionless units} All quantities in the text are reported in dimensionless units based on energy $\epsilon$, mass $m_s $, and distance $a_0$, parameters: $t(m_s a_0^2 / \epsilon)^{1/2} \rightarrow t$, $r / a_0 \rightarrow r$, and $k_B T / \epsilon \rightarrow T$.

\bibliography{ref}

\textbf{Acknowledgements}
The authors thank Liyan Qiao for useful discussions. This work was supported in part by the National Natural Science Foundation of China (No.: 12274110),  Natural Science Foundation of Zhejiang Province (No.: LHZSZ26A040002), and the Natural Sciences and Engineering Research Council of Canada.

\textbf{Author contributions}
J. C. and R. K. contributed at all stages of this work. J. H. performed part of simulations.

\textbf{Competing interests}: The authors declare no competing financial interests.

\end{document}